\documentclass[twocolumn,showpacs,preprintnumbers,amsmath,amssymb]{revtex4}

\usepackage{graphicx}
\usepackage{dcolumn}
\usepackage{bm}
\usepackage{amsmath}

\begin{document}

\title{Oscillations of Echo Amplitude in Glasses in a Magnetic Field\\
Induced by Nuclear Dipole-Dipole Interaction.}

\author{A.\,V. Shumilin}
\affiliation{A.\,F.\,Ioffe Institute, 194021 Saint Petersburg, Russia.}

\author{D.\,A. Parshin}

\affiliation{Saint Petersburg State Polytechnical University, 195251
Saint Petersburg, Russia }

\date{\today}

\begin{abstract}
The effect of a magnetic field on the dipole echo amplitude in
glasses (at temperatures of about $10$\,mK) induced by the
dipole-dipole interaction of nuclear spins has been theoretically
studied. It has been shown that a change in the positions of
nuclear spins as a result of tunneling and their
interaction with the external magnetic field $E_H$ lead to a nonmonotonic
magnetic field dependence of the dipole echo amplitude. The
approximation that the nuclear dipole-dipole interaction energy
$E_d$ is much smaller than the Zeeman energy $E_H$ has been found to be
valid in the experimentally important cases. It has been shown that
the dipole echo amplitude in this approximation may be described by
a simple universal analytic function independent of the microscopic
structure of the two-level systems. An excellent agreement of the
theory with the experimental data has been obtained without fitting
parameters (except for the unknown echo amplitude).
\end{abstract}



\maketitle

\section{Introduction}

Glasses at temperatures below $1$\,K are known to exhibit a number
of universal properties almost independent of their composition and
fundamentally different from the properties of similar crystals.
These properties are traditionally described within the model of
two-level systems (TLS's) \cite{AGVF}. One of these properties is
a two-pulse electric dipole echo in glasses, which is the delayed
response to two high-frequency electromagnetic pulses.

When a glass is subjected to two rf pulses with a frequency of about
$1$\,GHz separated by a time interval $\tau$ much longer than the
pulse length, one can observe a response in the glass polarization
at time $\tau$ after the second pulse. Especially interesting is the
pronounced nonmonotonic magnetic field dependence of the two-pulse
echo amplitude observed at temperatures of $10$\,mK in the absence
of paramagnetic centers in the glass \cite{Albasi}. The nature of
this interesting phenomenon remained unclear until works
\cite{quadr,glyc,wurger,diagram}, where this dependence was
associated with the presence of atoms with a nuclear quadrupole
moment in the glass.

Owing to the interaction of the nuclear magnetic moment with the
external magnetic field (Zeeman interaction) and the nuclear quadrupole
moment with the gradient of the internal electric field, the levels
of the TLS split into two nearly identical series of levels. The
characteristic energies of such a fine structure are $10^{-21}$\,erg
(corresponding to a frequency of about $100$\,kHz), which are much
lower than the typical energy of $50$\,mK of the TLS's specified by
the $1$\,GHz frequency of the excitation pulses. Since the internal
field gradient is different at different positions of the TLS's, the
fine level structure depends on the tunneling state of the TLS, which
ultimately leads to the nonmonotonic magnetic field dependence of
the dipole echo amplitude \cite{quadr,wurger,diagram}.

As a proof of this hypothesis, the results on measuring the
two-pulse echo amplitude in glycerol ($\rm C_3H_8O_3$) were
presented \cite{glyc}. In this experiment the echo amplitude dependence on
magnetic field increased by more than an order of magnitude under
the replacement of hydrogen with zero nuclear quadrupole moment by
deuterium (whose nuclear spin is 1 so deuterium has a small
quadrupole moment).

The oscillations of the echo amplitude in a magnetic field
were theoretically studied in
\cite{glyc,wurger,diagram,dresden,JOP}. In particular, the
qualitative explanation of the observed effect was given in
\cite{glyc,wurger,diagram}. In \cite{dresden,JOP}, numerical
calculations of the dipole echo amplitude in glycerol in the
magnetic field were performed and a good agreement with the
experimental data was obtained.

In our previous work \cite{jetp1}, we qualitatively and, in some
cases, quantitatively compared our analytic expressions for the echo
amplitude in glasses containing nuclear electric quadrupole moments
with the experimental data. However, the existence of small
oscillations of the dipole echo amplitude in glasses containing only
spherical nuclei (without quadrupole moments), like in nondeuterated
glycerol $\rm C_3H_8O_3$, remained unexplained until papers
\cite{dresden,JOP}. The presence of a small ($0.037\%$) natural
impurity of the $\rm ^{17}O$ isotope, which also has a nuclear
quadrupole moment, did not explain these oscillations.

Fleishmann \cite{dresden} suggested that the magnetic field
dependence of the echo amplitude in the case of non-deuterated $\rm
C_3H_8O_3$ may be caused by the dipole-dipole interaction of the
nuclear magnetic moments of hydrogen atoms (the spin and,
respectively, the magnetic moment of the primary isotopes of carbon
$^{12}$C and oxygen $^{16}$O are zero). This interaction
(along with the Zeeman interaction) induces a fine structure of the
levels of the TLS, which depends on the tunneling states of the
systems.

Bazrafshan et al. \cite{JOP} numerically calculated the echo
amplitude in deuterated glycerol $\rm C_3D_5H_3O_3$ (taking into
account all of the spins of hydrogen and the quadrupole moments of
deuterium). It was assumed that the tunneling of the two-level system is
the rotation of the glycerol molecule as a whole. Although the
theory was in a good agreement with the experiment in high fields and
in semiquantitative agreement in low magnetic fields (when the
dipole-dipole interaction is responsible for the effect), the
assumption of the rotational character of tunneling was not
justified.

This work is aimed at obtaining the analytical results for the
magnetic field dependence of the dipole echo amplitude in the
absence of nonspherical nuclei in the glass, so that only the
dipole-dipole interaction of the nuclear spins is taken into
account, and at comparing the results with the experimental data on
non-deuterated glycerol.

\section{General theory of the magnetic field dependence of the echo amplitude}

Parshin \cite{diagram} derived a general formula specifying the
dipole echo amplitude in an arbitrary system, which consists of two
identical sets of levels separated by an energy gap much larger than
the splitting within the sets
\begin{multline}
\label{gen1} P_{\rm echo}\propto-\frac{\displaystyle
i}{\displaystyle N}
V_1V_2^2  \cdot{} \\
{}\cdot \sum_{n,k} e^{i(E_n-E_k)\tau/\hbar}
 \left| \sum_m \alpha_{nm}^{(12)}
\alpha_{km}^{*(12)}e^{iE_m\tau/\hbar}\right|^2 .
\end{multline}
Here, $P_{\rm echo}$ is the dipole echo amplitude; $V_1$ and $V_2$
are the amplitudes of the first and second exciting electric pulses,
respectively; $E_n$ are the fine-structure energy levels of the TLS,
$N$ is the total number of this levels, and $\alpha_{nm}^{(12)}$ is
the matrix element for the transition between the n-th lower and
m-th higher levels of the split TLS during the action of the
excitation pulses.

To use this formula, we have to find the fine-structure energy
levels and the matrix elements of the transition in an arbitrary
magnetic field. For that, let us write the Hamiltonian of the system
including, first, the tunneling of the TLS atoms and, second, the
energies of the nuclear spins, which form the fine level structure:
\begin{equation}
\label{H0} \widehat{H}_{\rm tot} = \widehat{H}_{\rm tls}\otimes
\widehat{1}_J + \widehat{1}_{\rm tls}\otimes \widehat{W}_J +
\sigma_{z,\rm tls}\otimes\widehat{V}_J + (\widehat{\bf d}_{\rm
tls}{\bf F})\otimes \widehat{1}_J.
\end{equation}
Here $\widehat{H}_{\rm tls}$ is the Hamiltonian of the TLS in the
coordinate representation (excluding the fine structure);
$\widehat{1}_{\rm tls}$ and $\widehat{1}_J$ are the identity
matrices in the spaces of the TLS ($2 \times 2$) and the nuclear
spin projections, respectively; $\sigma_{z,\rm tls}$ is the Pauli
matrix for the two-level system; $\widehat{\bf d}_{\rm tls} = {\bf
d}\sigma_{z, {\rm tls}}$ is the operator of the electric dipole
moment $\bf d$ of the TLS; and $\bf F$ is the external (excitation)
electric field.

The spin operators $\widehat{W}_J$ and $\widehat{V}_J$ are,
respectively, the symmetric and antisymmetric (with respect to the
TLS displacement) parts of the spin Hamiltonian
\begin{equation}
\label{HJ} \widehat{H}_J^{(1,2)} = \sum_i
\widehat{\boldsymbol{\mu}}_i {\bf H} + \frac{1}{2}\sum_{ij} \left[
\frac{\widehat{\boldsymbol{\mu}}_i
\widehat{\boldsymbol{\mu}}_j}{r_{ij}^3} -
3\frac{(\widehat{\boldsymbol{\mu}}_i {\bf
r}_{ij})(\widehat{\boldsymbol{\mu}}_j{\bf r}_{ij})}{r_{ij}^5}
\right].
\end{equation}
Here $\widehat{\boldsymbol{\mu}}_i = \mu_i\widehat{{\bf J}}_i/J_i$
is the operator of the magnetic moment of the i-th nuclei,
$\widehat{{\bf J}}_i$ is the operator of its spin, and $H$ is the
external magnetic field. The radius vector $r_{ij} \equiv
r_{ij}^{(1,2)}$ from the i-th to the j-th nuclei may depend on the
TLS position $(1)$ or $(2)$. The summation is performed over all
tunneling nuclei. Consequently,
\begin{equation}\label{WV}
  \widehat{W}_J = \frac{\widehat{H}_J^{(1)} + \widehat{H}_J^{(2)}}{2} ,\quad
  \widehat{V}_J = \frac{\widehat{H}_J^{(1)} - \widehat{H}_J^{(2)}}{2} .
\end{equation}

Note that the first term of spin Hamiltonian (\ref{HJ}), which is
associated with the external magnetic field (the Zeeman part), does
not include vectors ${\bf r}_{ij}$ that change under tunneling of
the TLS and, therefore, does not contribute to the antisymmetric
$\widehat{V}_J$ part of the Hamiltonian. The dipole-dipole
interaction (the second term of Hamiltonian (\ref{HJ})) enters both
operators $\widehat{W}_J$ and $\widehat{V}_J$.

To find the quantities associated with the fine level structure, let us
rewrite Hamiltonian (\ref{H0}) in the basis consisting of the
stationary wavefunctions of the bottom and top levels of the TLS and
the eigenfunctions of the operator $\widehat{W}_J$  for the spin
variables:
\begin{multline}
\label{dHam4} \widehat{H}_{\rm tot} = \frac{1}{2} \left(
\begin{array}{cc}
E& 0\\
0& -E \\
\end{array}
\right) \otimes \widehat{1}_J +
\widehat{1}_\sigma\otimes\widehat{\widetilde{W}}_J +{}\\
{}+ (\mbox{\bf F}\cdot\mbox{\bf d})\,\frac{1}{E}\, \left(
\begin{array}{lc}
\Delta & \Delta_0\\
\Delta_0 &- \Delta \\
\end{array}
\right) \otimes\widehat{1}_J
+{}\\
{}+ \frac{1}{E}\, \left(
\begin{array}{lc}
\Delta & \Delta_0\\
\Delta_0 &- \Delta \\
\end{array}\right)\otimes \widehat{\widetilde{V}}_J
.
\end{multline}
Here, $\Delta$ is the difference between the minima of the
double-well potential, $\Delta_0$ is the tunneling amplitude of the
initial (unsplit) TLS, and $E =\sqrt{\Delta^2+\Delta_0^2} $ is the
total energy of the TLS (without the inclusion of the fine-structure
effects). The tilded operators $\widehat{\widetilde{W}}_J$ and
$\widehat{\widetilde{V}}_J$ correspond to the choice of the
eigenfunctions of $\widehat{W}_J$ as the basis. In this basis,
$\widehat{\widetilde{W}}_J$ is a diagonal matrix.

We assume that the characteristic tunneling lengths of the nuclei
are much shorter than the characteristic interatomic distances and,
consequently, $\widehat{\widetilde{V}}_J$ is much smaller than
$\widehat{\widetilde{W}}_J$ and may be taken into account
perturbatively. Retaining only the terms proportional to
$\widehat{\widetilde{V}}_J^2$ in the transition matrix elements, we
obtain the dipole echo amplitude (compare with, e.g., \cite{diagram})
\begin{multline}
P_{\rm echo} \propto   \left(\frac{\Delta_0}{E}\right)^4 \times {}\\
{}\times \left[ 1 - \frac{64}{N} \left(\frac{\Delta}{E}\right)^2
\sum_{n,m>n}
 \left|(\widetilde{V}_J)_{nm}\right|^2
\frac{\sin^4\left(\varepsilon_{nm}\tau/2\hbar\right)}{\varepsilon_{nm}^2}
\right]. \label{gendip}
\end{multline}
Here,
$\varepsilon_{nm}=\left(\widetilde{W}_J\right)_{nn}-\left(\widetilde{W}_J\right)_{mm}
\equiv E_n - E_m$ is the difference between the fine-structure
levels of the TLS.

Expression (\ref{gendip}) allows one to calculate the echo amplitude
from one TLS. To find the observed dipole echo amplitude of the
entire sample, it is necessary to perform the summation over all
two-level systems.

To obtain the expression for the distance between the fine-structure
levels $\varepsilon_{nm}$, we have to diagonalize the matrix
$\widehat{W}_J$, which cannot be done analytically even in the
simplest case of two interacting spins $1/2$ (a $4 \times 4$
matrix). Thus, to go further, we need to perform some
simplifications.

The nonmonotonic magnetic field dependence of the dipole echo
amplitude in $\rm C_3H_8O_3$ was experimentally observed in the
magnetic fields $H \simeq 50$\,G, which corresponds to the Zeeman
energy $E_H \simeq 0.7 \times 10^{-21}$\,erg. The energy of the
dipole-dipole interaction between two hydrogen nuclei at a
distance of $1.9\,$\,\AA  ~(the minimum distance between hydrogen atoms
in glycerol) may be estimated as $E_d \simeq 1.5 \times
10^{-22}$\,erg, which is almost an order of magnitude smaller than
the Zeeman energy.

\section{Weak dipole-dipole interaction approximation}

In the $E_d \ll E_H$ approximation, we neglect the dipole-dipole
interaction in the expression for $\widehat{W}_J$. However, this
interaction should be retained in $\widehat{V}_J$, which does not
contain terms of the order of $E_H$.

In this approximation, the eigenfunctions of $\widehat{W}_J$ are the
states with definite projections of all nuclear spins on the
direction of the magnetic field. In this case, only the matrix
elements $(\widetilde{V}_J)_{nm}$ of the transition with a change of
no more then two spin projections are nonzero. This allows us to
reduce the problem of an arbitrary number of dipole-dipole
interacting nuclear spins to the problem of one pair of spins. We
consider only the pairs of identical spin $1/2$ nuclei. The operator
$\widehat{W}_J$ may be then written in the form
\begin{equation}
\label{Wq_pert} \widehat{W}_J = \mu {\bf H} (\widehat{\bf J}_1 +
\widehat{\bf J}_2),
\end{equation}
where $\mu$ is the nuclear magnetic moment and $\widehat{\bf J}_1$
and $\widehat{\bf J}_2$ are the spin operators of the interacting
nuclei. The matrix form of $\widehat{W}_J$ is
\begin{equation}
\label{WJ_matr} \widehat{W}_J = \left(
\begin{array}{cccc}
2\mu H & 0 & 0 & 0 \\
0 & 0 & 0 & 0 \\
0 & 0 & 0 & 0 \\
0 & 0 & 0 & -2\mu H
\end{array}
\right).
\end{equation}
It is easy to see that, in this case, there are four energy
differences $\varepsilon_{nm}$ equal to $2\mu H$ and one difference
$\varepsilon_{14} = 4\mu H$ ($\varepsilon_{23}$ is zero and does not
contribute to the echo amplitude).

To express the echo amplitude explicitly with the use of Eq.
(\ref{gendip}), we need to calculate also the average squares of the
matrix elements $|(\widetilde{V}_J )_{nm}|^2$. Taking into account
the smallness of the tunneling length, the
$\widehat{\widetilde{V}}_J$ operator in the case of two nuclei can
be represented as
\begin{equation}
\label{Ed1} \widehat{\widetilde{V}}_J = \delta {\bf r}_{12} \cdot
\frac{\partial}{\partial {\bf r}_{12}}\left[
\frac{\widehat{\boldsymbol{\mu}}_1
\widehat{\boldsymbol{\mu}}_2}{r_{12}^3} -
3\frac{(\widehat{\boldsymbol\mu}_1{\bf
r}_{12})(\widehat{\boldsymbol\mu}_2 {\bf r}_{12})}{r_{12}^5}
\right].
\end{equation}
Expressing the gradient explicitly and taking out $\delta r_{12}$,
$r_{12}$ and $\widehat{\mu}_{1,2}$ as the common factors, we may
rewrite Eq. (\ref{Ed1}) in the form
\begin{equation}
\label{Ed12} \widehat{\widetilde{V}}_J = \mu^2 \frac{\delta
r_{12}}{r_{12}^4}\widehat{\sigma}_{1,\alpha}
\widehat{\sigma}_{2,\beta} N_{\alpha\beta}({\bf n}_{r}, {\bf
n}_{\delta r}).
\end{equation}
Here, the Greek subscripts numerate the Cartesian coordinates;
$\widehat{\sigma}_{1,\alpha}$ and $\widehat{\sigma}_{2,\beta}$ are
the Pauli matrices for the first and second spin, respectively;
$N_{\alpha\beta}({\bf n}_r, {\bf n}_{\delta r})$ is the symmetric
second-rank tensor, which depends merely on the directions of the
vectors $\bf r$ (${\bf n}_r = {\bf r}/r$) and $\delta {\bf r}$
(${\bf n}_{\delta r} = \delta {\bf r}/\delta r)$.

To find the sought expression for $|(\widetilde{V}_J)_{nm}|^2$, one
has to express one matrix element from Eq. (\ref{Ed12}) and average
its absolute value square over the directions of the vectors ${\bf
n}_r$ and ${\bf n}_{\delta r}$ (it is easy to see that the averaging
affects only the tensor $N_{\alpha\beta}$). For this purpose, we
introduce a new numeration of the elements of
$\widehat{\widetilde{V}}_J$: $(\widetilde{V}_J)_{nm} =
(\widetilde{V}_J)_{(pr)(qs)}$. Here, the subscripts $p, q = \pm 1$ and
$r, s = \pm 1$ numerate the states of the first and second spins,
respectively. Correspondingly, $n = (pr)$ and $m = (qs)$. In this
notation,
\begin{multline}
\label{VJd}
|(\widetilde{V_J})_{(pr)(qs)}|^2 = \mu^4\frac{\delta r^2}{r^8} \times {} \\
{} \times
\sigma_{1,\alpha,pq}\sigma_{2,\beta,rs}\sigma_{1,\xi,pq}^*\sigma_{2,\zeta,rs}^*
\left< N_{\alpha\beta}N_{\xi\zeta} \right>.
\end{multline}
Here, $\sigma_{1,\alpha,pq}$ is the $pq$  element of the Pauli
matrix $\sigma_\alpha$ acting on the wavefunctions of the first
spin. $\sigma_{2,\beta,rs}$ is the same for the second spin.

\begin{figure}[htbp]
    \centering
        \includegraphics[width=0.35\textwidth]{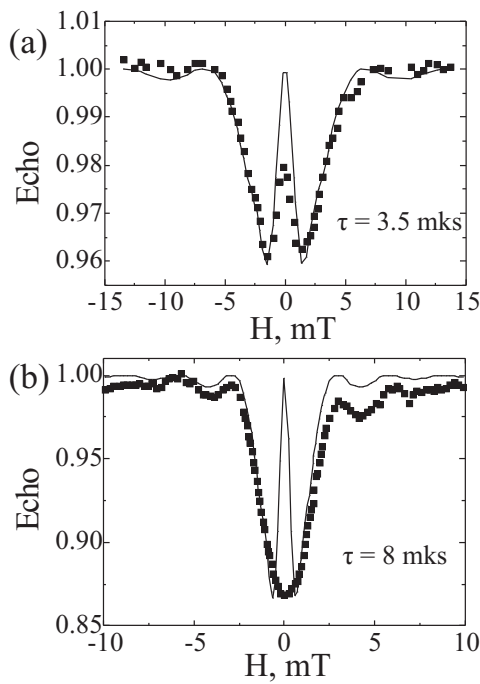}
        \caption{Experimental data on glycerol \cite{glyc,diplom} at (a) $\tau =3.5\,\mu s$ and (b)
        $8\,\mu s$ in comparison with Eq. (\ref{pertecho}) with $\mu$ taken equal to the proton
        magnetic moment.}
    \label{fig:dip}
\end{figure}

 The isotropy of the glass allows us to express the
symmetric tensor  $\left< N_{\alpha\beta}N_{\xi\zeta} \right>$ in
terms of two constants, $\left< N_{\alpha\beta}N_{\xi\zeta} \right>
= A\delta_{\alpha\beta}\delta_{\xi\zeta} +
B(\delta_{\alpha\xi}\delta_{\beta\zeta} +
\delta_{\alpha\zeta}\delta_{\beta\xi})$. However, it can be shown by
direct calculation that the quantity
$\sigma_{1,\alpha,pq}\sigma_{2,\beta,rs}\sigma_{1,\xi,pq}^*\sigma_{2,\zeta,rs}^*
\delta_{\alpha\beta}\delta_{\xi\zeta} =
|\boldsymbol{\sigma}_{1,pq}\cdot\boldsymbol{\sigma}_{2,rs}|^2$ ,
which corresponds to the constant $A$, includes only the diagonal
elements and the elements corresponding to the transition between
the degenerate second and third levels of $\widehat{W}_J$. Neither
of them contributes to the echo amplitude and, consequently, all
significant matrix elements of $\widehat{\widetilde{V}}_J$ may be
expressed in terms of a single constant $B$.

Performing the calculations, we arrive at the final expression for
the dipole echo amplitude in the form
\begin{equation}
\label{pertecho} P_{\rm echo} \propto 1 - C \left[ \frac{\sin^4(\mu
H\tau/\hbar)}{(\mu H \tau)^2} + \frac{\sin^4(2\mu
H\tau/\hbar)}{4(\mu H \tau)^2 } \right] ,
\end{equation}
where $C$ is a constant independent of the magnetic field.

Note that expression (\ref{pertecho}) was derived without any
assumptions about the relation between ${\bf n}_r$ and ${\bf
n}_{\delta r}$ and, consequently, is absolutely independent of the
tunneling geometry of the atoms, that is, of the microscopic structure of
the TLS.

Let us compare Eq. (\ref{pertecho}) with the measurements
\cite{glyc,diplom} of the magnetic field dependence of the dipole
echo amplitude in glycerol (see figure). Clearly, the analytical
curve with $\mu$ equal to the magnetic moment of a proton describes
the experimental data sufficiently well without any fitting
parameters (except for the constant $C$, which specifies the
vertical scale) until the region of small magnetic fields ($\le
1$\,mT), where the conditions $E_H \ll Ed$ seemingly breaks.

\subsection{Applicability Conditions}

The nonmonotonic magnetic field dependence of the dipole echo
amplitude is determined by the term
$\sin^4(\varepsilon_{nm}\tau/2\hbar)$ in Eq. (\ref{gendip}). This
corresponds to the characteristic energies $\varepsilon_{nm} \simeq
 2\hbar/\tau$. In the case of $E_d \ll 2\hbar/\tau$ (or, equivalently, $\tau \ll 2\hbar/E_d$), we
may expect that the condition $E_d \ll E_H$ holds for the dominant
part of the dependence. Otherwise, the exact inclusion of the
dipole-dipole interaction is required; in this case, in particular,
the problem cannot be reduced to the problem of two tunneling
nuclei. This agrees well with the fact that the theory better
describes the experimental data for $\tau = 3.5\,\mu s$ than for
$\tau = 8\,\mu s$ (see figure) in the low-field region.

Note that the horizontal scale of the magnetic field dependence of
the echo amplitude in the case under consideration is determined by
the time $\tau$. Thus, the curve $P_{\rm echo}(H)$ should scale in
the horizontal axis as $1/\tau$ with a change in the time interval
between the pulses. In the opposite case of $\tau \gtrsim 2\hbar/
E_d$, the horizontal scale of the dependence is determined by the
relation $E_d \simeq E_H$ and is independent of the time interval
between the pulses. This may be used as an experimental criterion of
the smallness of the dipole-dipole interaction.

To conclude, in the experimentally observed case of $\tau \ll
2\hbar/E_d$, the magnetic field dependence of the dipole echo amplitude
induced by the dipole-dipole interaction of the nuclei of the same
kind has a universal character independent of the microscopic
structure of the TLS. It is well described by a simple analytic expression. A
good agreement of the theory and experiment on the dipole echo in
nondeuterated glycerol was obtained without any fitting parameter
(except for the echo amplitude scale). At larger values of the time
interval between the excitation pulses, $\tau \gtrsim 2\hbar/E_d$,
the universal form of the dependence breaks, which may allow one to
obtain information on the microscopic structure of the two-level
systems with the use of numerical analysis \cite{ltc08}.

We are grateful to A. Fleishmann and M. Bazrafshan for fruitful
discussions.

A.V.S. thanks the St.~Petersburg administration for
support in 2008, candidate project no. 2.4/4-05/103.

\end{document}